\pdfoutput=1
\documentclass[11pt,twoside,a4paper,cmspaper,final,collab]{cms-tdr}
\def\svnVersion{753b3c1}\def\svnDate{2022/03/11}\def\cmsCernNoTag{CERN-EP-2021-014}\def\cmsCernDate{\today}\def\cmsMessage{Published in Physical Review Letters as \href{http://dx.doi.org/10.1103/PhysRevLett.128.122301}{\doi{10.1103/PhysRevLett.128.122301}.}}
\begin{document}\cmsNoteHeader{HIN-19-006}

\ifthenelse{\boolean{cms@external}}{\providecommand{\suppMaterial}{the supplemental material~\cite{supp}\xspace}}{\providecommand{\suppMaterial}{Appendix~\ref{app:suppMat}\nocite{supp}\xspace}}

\ifthenelse{\boolean{cms@external}}{\providecommand{\suppRef}{1 in the supplemental material\xspace}}{\providecommand{\suppRef}{\ref{fig:trkpt_data} in the Appendix\xspace}}

\newcommand {\roots}    {\ensuremath{\sqrt{s}}}
\newcommand{\doublequotes}[1]{``#1''}

\newcommand {\zjet}{Z+jet\xspace}

\newcommand {\pp}    {\ensuremath{\Pp\Pp}\xspace}
\newcommand {\PbPb}  {\mbox{PbPb}\xspace}

\newcommand {\ptz}    {\ensuremath{\pt^\text{\PZ}}\xspace}
\newcommand {\ptt}    {\ensuremath{\pt^\text{trk}}\xspace}

\newcommand {\pttvec}    {\ensuremath{\ptvec^\text{trk}}\xspace}
\newcommand {\ptzvec}    {\ensuremath{\ptvec^\text{\PZ}}\xspace}

\newcommand {\etat}    {\ensuremath{\eta^\text{trk}}\xspace}
\newcommand {\deta}     {\ensuremath{\Delta\eta}\xspace}
\newcommand {\dphi}     {\ensuremath{\Delta\phi}\xspace}

\newcommand {\phiz}    {\ensuremath{\phi^\text{\PZ}}\xspace}
\newcommand {\phit}    {\ensuremath{\phi^\text{trk}}\xspace}

\newcommand {\energyHF}    {\ensuremath{E^\text{HF}}\xspace}
\newcommand {\energyHFZ}    {\ensuremath{E^\text{HF, \PZ}}\xspace}
\newcommand {\energyHFZpp}    {\ensuremath{\langle E^\text{HF, \PZ, \pp}\rangle}\xspace}

\newcommand{\dphitz}{\ensuremath{\Delta\phi_{\mathrm{trk,\PZ}}}\xspace}
\newcommand{\xizt}  {\ensuremath{\xi^\mathrm{trk,\PZ}_{\mathrm{T}}}\xspace}

\newcommand{\NtrkZNZ}{\ensuremath{ \mathrm{N_{trk,\PZ}} / \mathrm{N_{\PZ}} }\xspace}

\newcommand{\dNdphitz}{\ensuremath{1/\mathrm{N_{\PZ}}\, \rd\mathrm{N_\text{{trk,\PZ}}}/\rd\dphitz}\xspace}
\newcommand{\dNxizt}{\ensuremath{1/\mathrm{N_{\PZ}}\, \rd\mathrm{N_\text{{trk,\PZ}}}/\rd\xizt}\xspace}
\newcommand{\dNptt}{\ensuremath{1/\mathrm{N_{\PZ}}\, \rd\mathrm{N_\text{{trk,\PZ}}}/\rd\ptt}\xspace}

\newcommand{\EMBEDDED} {\text{embedded}\xspace}
\newcommand{\scet} {{SCET}\xspace}
\newcommand{\scetg} {\ensuremath{{\mathrm{{\scet}_G}}}\xspace}

\cmsNoteHeader{HIN-19-006}

\title{Using \texorpdfstring{\PZ}{Z}  boson events to study parton-medium interactions in \texorpdfstring{\PbPb}{PbPb} collisions}
\date{\today}

\abstract{
The spectra measurements of charged hadrons produced in the shower of a parton originating in the same hard scattering with a leptonically decaying \PZ boson, are reported in lead-lead nuclei (\PbPb) and proton-proton (\pp) collisions at a nucleon-nucleon center-of-mass energy of 5.02\TeV. Both \PbPb and \pp data sets are recorded by the CMS experiment at the LHC, and correspond to an integrated luminosity of 1.7\nbinv and 320\pbinv, respectively. Hadronic collision data with one reconstructed \PZ boson candidate with the transverse momentum $\pt > 30\GeVc$ are analyzed. The \PZ boson constrains the initial energy and direction of the associated parton. In heavy ion events, azimuthal angular distributions of charged hadrons with respect to the direction of a \PZ boson are sensitive to modifications of the in-medium parton shower and medium response. Compared to reference data from \pp interactions, the results for central \PbPb collisions indicate a modification of the angular correlations. The measurements of the fragmentation functions and \pt spectra of charged particles in \PZ boson events, which are sensitive to medium modifications of the parton shower longitudinal structure, are also reported. Significant modifications in central \PbPb events compared to \pp reference data are also found for these observables.}

\hypersetup{
pdfauthor={CMS Collaboration},
pdftitle={Using Z boson events to study parton-medium interactions in PbPb collisions},
pdfsubject={CMS},
pdfkeywords={CMS, quark-gluon plasma, Z boson, relativistic heavy ion collisions}}

\maketitle 

In relativistic heavy ion collisions, quantum chromodynamics (QCD) predicts that a state of deconfined quarks and gluons, known as quark-gluon plasma (QGP), can be formed~\cite{PhysRevLett.34.1353,Karsch:1995sy}. 
Parton scatterings with large momentum transfer, which occur very early in the collision compared to the timescale of QGP formation, can act as tomographic probes of the plasma~\cite{Appel:1985dq}. The outgoing partons interact strongly with the QGP and lose energy~\cite{Blaizot:1986ma,Gyulassy:1990ye}, resulting in showers with more particles of lower energy. This phenomenon, known as ``jet quenching'', has been observed through measurements of hadrons with high transverse momentum with respect to the beam direction (\pt)~\cite{Adams:2003kv,Back:2004ra,Adare:2008qa,Abelev:2012hxa,Aad:2015wga,Khachatryan:2016odn} and of jets~\cite{CMS:2021vui,Khachatryan:2016jfl,ALICE:2015mjv,ALICE:2019qyj,Adam:2015doa,Adamczyk:2016fqm,Chatrchyan:2011sx,ATLAS:2018gwx, ATLAS:2010isq}, both created by the fragmentation of energetic partons.

This Letter presents the measurement of charged hadrons from the shower of a parton (quark or gluon) produced in association with a \PZ boson in lead-lead nuclei (\PbPb) and proton-proton (\pp) collisions. Both \PbPb and \pp data sets are collected at a nucleon-nucleon center-of-mass energy $\sqrtsNN = 5.02\TeV$ and correspond to integrated luminosities of 1.7\nbinv and 320\pbinv, respectively. The advantage~\cite{Kartvelishvili:1995fr,Wang:1996yh,Kang:2017xnc} of measuring jets produced in the same hard scattering with an electroweak boson (\eg, photon, \PZ or \PW bosons) arises because these do not interact strongly with the QGP~\cite{Aad:2015lcb,Chatrchyan:2012vq,Chatrchyan:2012nt,Chatrchyan:2014csa}. The initial direction and energy of the associated parton that fragments into the jet, before any medium-induced energy loss happens, is determined, in the transverse plane, by the momentum of the electroweak boson (the `tag`), on average (\ie, the kinematic balance of the outgoing particles can be slightly distorted by processes that happen even in the absence of a QGP). There are several advantages to using a \PZ boson as a tag instead of a photon: minimal contributions from other background channels~\cite{Neufeld:2012df, Casalderrey-Solana:2015vaa, KunnawalkamElayavalli:2016ttl, Kang:2017xnc}, absence of irreducible background sources~\cite{Chatrchyan:2012vq, Chatrchyan:2012gt}, and smaller uncertainties arising from the experimental selection and identification of \PZ boson candidates. 

The goals of this measurement are the following: (i) to study the medium modification of the hadron momentum spectra coming from hard-scattered partons tagged by \PZ bosons~\cite{Chen:2020tbl,Kang:2017xnc,Ovanesyan:2011xy}, (ii) to reveal possible angular decorrelations between the unmodified \PZ boson direction and the charged hadrons because of \pt broadening originating from interactions of the parent parton with the medium~\cite{Chen:2017zte,Casalderrey-Solana:2014bpa}, and (iii) to study the possible effects of medium recoil in the angular correlation functions between the charged hadrons from the shower of a parton produced in association with a \PZ boson~\cite{Chen:2020tbl,Casalderrey-Solana:2016jvj,Ovanesyan:2011xy}. This analysis correlates \PZ bosons (reconstructed when decaying to pairs of electrons or muons) and charged-particle tracks in relative azimuthal angle ($\phi$). The \NtrkZNZ, the number of tracks normalized by the number of \PZ bosons, is measured as a function of the difference between the $\phi$ angle of the \PZ boson (\phiz) and the angles (\phit) of the other tracks reconstructed in the event, $\dphitz = \abs{\phit - \phiz}$. This Letter also presents measurements of the longitudinal momentum distribution of \PZ-tagged jet constituents, \ie, the jet fragmentation variable $\xizt = \ln{\bigl[-\abs{\ptzvec}^{2}/(\pttvec \cdot \ptzvec)\bigr]}$, where \ptzvec and \pttvec are the \pt vectors with respect to the beam direction of the \PZ boson and charged-particle track, respectively~\cite{Casalderrey-Solana:2015vaa}. These results are distinct from previous $\xi$ measurements~\cite{Chatrchyan:2014ava} in which the \ptzvec in the denominator is replaced by the \pt of a jet after it suffered medium-induced energy loss. They are complementary to photon-tagged measurements~\cite{Sirunyan:2018qec,ATLAS:2019dsv} (where effects were probed for partons with higher initial \pt) and to other \PZ-tagged measurements~\cite{ATLAS:2020wmg} (where different \ptz selections were used to test the sensitivity of energy loss processes to various initial \pt of the partons). 

The central feature of the CMS detector~\cite{Chatrchyan:2008zzk} is a superconducting solenoid of 6\unit{m} internal diameter, providing a magnetic field of 3.8\unit{T}. Within the solenoid volume are a silicon pixel and strip tracker, a lead tungstate crystal electromagnetic calorimeter (ECAL), and a brass and scintillator hadron calorimeter. Hadron forward (HF) calorimeters extend the pseudorapidity coverage up to $\abs{\eta}=5.2$. For \PbPb events, the HF signals are used to determine the degree of overlap (``centrality'') of the two colliding nuclei~\cite{Chatrchyan:2011sx}. Muons are measured in gas-ionization detectors located outside the solenoid. 

The event samples are selected in real time with dedicated lepton filters (``triggers")~\cite{Khachatryan:2016bia}, and offline by removing noncollision events~\cite{Khachatryan:2016odn}. The $\PZ \to \Pep\Pem$ events are triggered if one ECAL cluster has transverse energy greater than 20\GeV and $\abs{\eta}< 2.1$, while the $\PZ \to \PGmp\PGmm$ triggers require one muon of $\pt > 12\GeVc$ and $\abs{\eta}< 2.4$~\cite{Khachatryan:2016bia}. The average pileup (the mean of the number of additional collisions within the same bunch crossing) is 2 in \pp, and negligible in \PbPb collisions. For \PbPb collisions, the results are presented in four centrality intervals, 70--90, 50--70, 30--50, and 0--30\%. The centrality measurement is based on percentiles of the distributions of the total energy deposited in the HF calorimeters, which corresponds to the fraction of the total inelastic hadronic cross section, starting at 0\% for the most central collisions~\cite{Chatrchyan:2011sx}.

The \PYTHIA~8.212~\cite{Sjostrand:2014zea} Monte Carlo (MC) event generator with the underlying event (UE) tune CP5~\cite{Sirunyan:2019dfx}, and \MGvATNLO~8.212~\cite{Alwall:2014hca} next-to-leading order (NLO) program (interfaced with \PYTHIA) are used to simulate \zjet signal events. In the \PbPb case, \doublequotes{\EMBEDDED} samples are created by overlapping \PYTHIA and \MGvATNLO signal events with minimum bias (MB) heavy ion events generated with the {\HYDJET}~1.9 MC event generator~\cite{Lokhtin:2005px}. The generated embedded events are propagated through the CMS apparatus using the \GEANTfour toolkit~\cite{geant4}.  These MC samples are used to evaluate reconstruction and selection efficiencies, calibrations, and to study the background. All evaluations and studies are carried separately for the \pp and \PbPb data. 

Electrons are identified as ECAL superclusters~\cite{Khachatryan:2015hwa} matched in position and energy to tracks reconstructed in the tracker, using the particle-flow algorithm~\cite{Sirunyan:2017ulk}. They must have $\pt > 20\GeVc$ and their supercluster must be within the acceptance of the trigger, $\abs{\eta}< 2.1$. Muons are selected by requiring reconstructed track segments in at least two muon detector planes and a good-quality fit when connecting them to tracker segments~\cite{Chatrchyan:2012xi}. For both \pp and \PbPb data, the muons are required to have $\pt > 20\GeVc$ and they must fall within the acceptance of the muon detectors, $\abs{\eta}<2.4$.

The track reconstruction used in \pp and \PbPb collisions is described in Ref.~\cite{Chatrchyan:2014fea}. Corrections for the tracking efficiency, detector acceptance, and misreconstruction rate are obtained following the procedure in Ref.~\cite{Khachatryan:2016odn}. Additional corrections are applied to account for a difference in tracking efficiency ($\sim$1\%), from a different particle density, seen between \HYDJET and \EMBEDDED \MGvATNLO samples. The selection criteria are the same as in Ref.~\cite{Khachatryan:2016odn} for both the \pp and \PbPb data.

The \PZ candidates are identified using an electron or muon pair, with a reconstructed invariant mass in the interval 60--120\GeVcc and $\ptz>30\GeVc$. After all selections, there are $\sim$5K (23K) \PZ boson events in the \PbPb (\pp) data. Electron and muon pairs are corrected for losses in acceptance and efficiency during reconstruction, and identification and trigger selections~\cite{Chatrchyan:2012xi,Khachatryan:2015hwa}. Each \PZ candidate is paired with all tracks in the same event that pass the $\ptt > 1\GeVc$ and $\abs{\etat} < 2.4$ selections. To avoid including the tracks of the \PZ candidate decay products, each track used in the correlations is required to fall outside a cone radius (defined as $\sqrt{\smash[b]{{\left( \deta \right )}^2 + {\left( \dphi \right)}^2}}$ ) of 0.02 (the smallest value for which no significant contamination is observed) around the direction of a lepton from the \PZ decay. Intermediate results, corrected for lepton efficiencies only, are obtained separately for \PZ candidates reconstructed from oppositely-charged electron or muon pairs. The residual ($<$3\%) contamination from QCD jet physics processes is estimated using same-charge lepton pairs, whose distributions are subtracted from those of opposite-charge leptons for each of the two decay channels.

Combinatorial background originating from tracks from the UE in \PbPb collisions is subtracted to obtain the correlation between the \PZ boson candidate and all tracks coming from the shower of a parton produced in the same nucleon-nucleon interaction. This background is estimated from data with an event mixing procedure~\cite{Sirunyan:2018qec,Sirunyan:2018ncy}, where the \PZ candidate is paired with tracks found in events chosen randomly from an MB \PbPb data set with similar event characteristics (\ie, similar energy deposited in the HF, and interaction vertex $z$ position). Events are split into bins of total HF calorimeter energy, \energyHF. 
To ensure that the \PZ boson and MB events have the same size UE, an event with a \PZ boson candidate and with \energyHFZ is mixed with MB events in the \energyHF bin containing events with HF energy equal to $\energyHFZ-\energyHFZpp$. The quantity \energyHFZpp is the average of \energyHF over events in the \pp data selected such that they contain a \PZ boson but no additional \pp pileup. The bin size is chosen such that it is narrow enough to have `closure' during MC tests (\ie, an agreement between generated charged particle yields from \zjet events, and reconstructed track yields from background-subtracted events). For the events within a given \energyHF bin, the variation in the number of UE tracks before subtraction can be much larger than the number of tracks after subtraction. In order to reflect this statistical effect of the UE, the statistical uncertainties of the \PbPb distributions are calculated using the bootstrap method~\cite{efron1979}. Dedicated tests based on control samples in data show that the UE produced by a \PZ boson process in a \PbPb collision is the same as in a \pp collision, within the statistical uncertainties of the present samples. It was checked that the results obtained using information only from the $\eta<0$ or only from the $\eta>0$ regions of the HF calorimeters are consistent with the main result. The UE subtraction procedure was validated by performing the whole analysis on MC embedded samples. The results obtained using the generated particles versus using the reconstructed (after UE subtraction) particles were compared, and any discrepancy was included in the systematic uncertainties.

Several variations in the analysis are considered in order to account for the uncertainties related to the tracking efficiency and corrections, lepton efficiency and energy scale, as well as \pp pileup and \PbPb background subtraction. No significant differences are observed in the results obtained with electron and muon pairs separately therefore, uncertainties are quoted after combining the two. With the exception of the lepton energy scale and efficiencies, there are no assumed correlations between the \pp and \PbPb uncertainties. Unless noted otherwise, the systematic uncertainties are evaluated as the differences between the final results and results obtained with varied settings. In the following, we list the variations considered, and provide in \suppMaterial the numerical values for the average uncertainties corresponding to the most extreme cases, \ie, the \pp and most central 0--30\% \PbPb collisions. 

{\tolerance=1600
The uncertainty related to the tracking efficiency is estimated as the difference in the track reconstruction efficiency between data and simulation~\cite{Khachatryan:2016odn}. The uncertainty related to the correction for the observed efficiency difference between \HYDJET and \EMBEDDED \MGvATNLO samples is obtained by comparing the corrections obtained from \MGvATNLO and \PYTHIA \EMBEDDED samples. Lepton efficiencies are varied by the uncertainty in their data-to-MC differences obtained using the \doublequotes{tag-and-probe} method~\cite{CMS:2011aa}. To assess the uncertainty related to the lepton energy scale corrections, the \pt of leptons is shifted by their energy correction uncertainties. No corrections are applied to remove the residual pileup effects in \pp data. Nominal distributions (no requirement on pileup) are compared to those from events without pileup, \ie, events with only one interaction vertex. The uncertainty in the event-mixing procedure is obtained by repeating the procedure after shifting the \energyHFZpp by 5\%, the maximum difference in the HF response between the \PbPb and \pp data-taking periods. Because the difference in the HF response between the beginning and end of the \PbPb run was found to be negligible ($<1$\%), no additional uncertainty was assigned.
\par} 

Three theoretical calculations are compared to the results; they use the same kinematic selection as data and incorporate the phenomenon of jet quenching, and differ just in their treatment of the medium response to the passing parton: \scetg~\cite{Li:2019dre,Ovanesyan:2011xy,Chien:2015vja}, which does not consider any medium response to jet propagation; Hybrid~\cite{Casalderrey-Solana:2014bpa,Casalderrey-Solana:2016jvj}, which considers the effects of a `wake', induced by the jet as it passes through and interacts with the QCD medium; and CoLBT~\cite{Chen:2020tbl,Yang:2021qtl} in which the quenched jet energy feeds into the hydrodynamic evolution. 

\begin{figure*}[hbt!]
  \centering
      \includegraphics[width=1.0\textwidth]{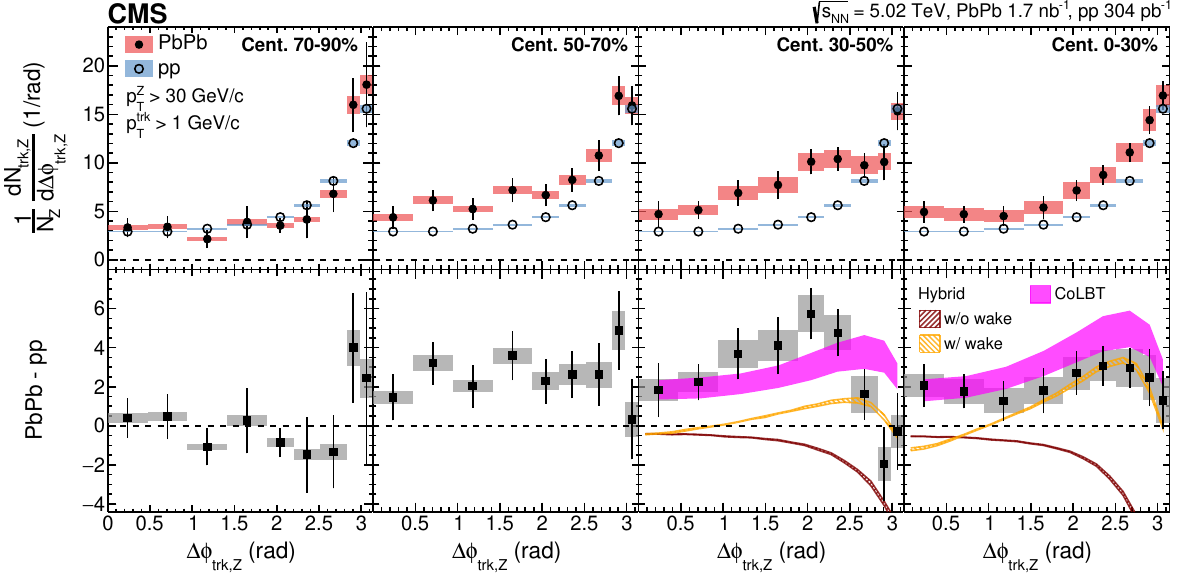}
      \caption{Upper: distributions of \dphitz in \pp collisions compared to \PbPb collisions (left to right) in the 70--90 (left), 50--70, 30--50, and 0--30\% (right) centrality intervals. Lower: difference between the \PbPb and \pp distributions. The vertical bars and shaded boxes represent the statistical and systematic uncertainties, respectively. Several model calculations are added for comparison: Hybrid~\cite{Casalderrey-Solana:2016jvj} and CoLBT~\cite{Chen:2020tbl,Yang:2021qtl}.}
      \label{fig:dphi_data}
\end{figure*}

Figure~\ref{fig:dphi_data} shows \dNdphitz, \ie, the distributions of the $\phi$ angle difference between charged particles and \PZ bosons, normalized by the number of \PZ bosons in each data set (and for the \PbPb case, in each centrality interval). This type of angular correlation function could reveal medium-induced modification of the away-side ($\dphitz\sim\pi$) jet constituents, and effects of the medium response (\ie, modification of the medium induced by the jet traversing through), over all $\dphitz$. Different pairs of data sets were compared using $\chi^2$-tests. With a $p$-value cutoff of 0.05, the tests show that the 0--30\% \PbPb distribution is compatible (\ie, statistically indistinguishable) with all data sets except the most peripheral one. In turn, the \pp distribution is found to be compatible only with the 70--90\% PbPb data set. The distributions in both \pp and \PbPb collisions are peaked at $\dphitz \sim\pi$, which is the signature of an away-side jet emitted back-to-back with the \PZ boson. None of the \PbPb or \pp distributions reach zero even in the $\dphitz \sim 0$ region, around the tag \PZ boson, in its direction of propagation. This happens even if (i) the random combinations from UE (between the \PZ candidates and tracks produced in nucleon-nucleon interactions that are independent of the \zjet process) have been removed using the event-mixing procedure, and (ii) the \PZ boson does not interact strongly with the medium in \PbPb collisions while traversing it, and iii) it is not produced during the fragmentation of a parton in \PbPb or \pp collisions (processes that could create more particles in the direction of propagation of the \PZ boson). 

The difference in the number of associated particles, between the \PbPb and \pp results, is also shown in Fig.~\ref{fig:dphi_data}. A $\chi^2$-test was done to assess the hypothesis that the excess observed is $\dphitz$-dependent: with the current precision of the measurement this hypothesis is rejected at the 95\% confidence level (\ie, the data are consistent with an increase of the yield that is independent of $\dphitz$). The excess observed in all bins except the most peripheral (\ie, the most \pp-like) could be caused by medium response, where the traversing jet excites the medium around it. Another possible contribution to the excess could be medium modifications of partons originating from the same nucleon-nucleon collision as the \zjet process, but from a different parton-parton interaction, and which would add a flat contribution over the entire $\dphitz$ range~\cite{Yang:2021qtl}. The comparison with the CoLBT and the Hybrid (with and without wake) models supports these scenarios, although the Hybrid model fails to reproduce the magnitude of the difference between \pp and \PbPb collisions, in particular in the $\dphitz \sim 0$ region.

\begin{figure*}[hbt!]
   \centering
      \includegraphics[width=1.0\textwidth]{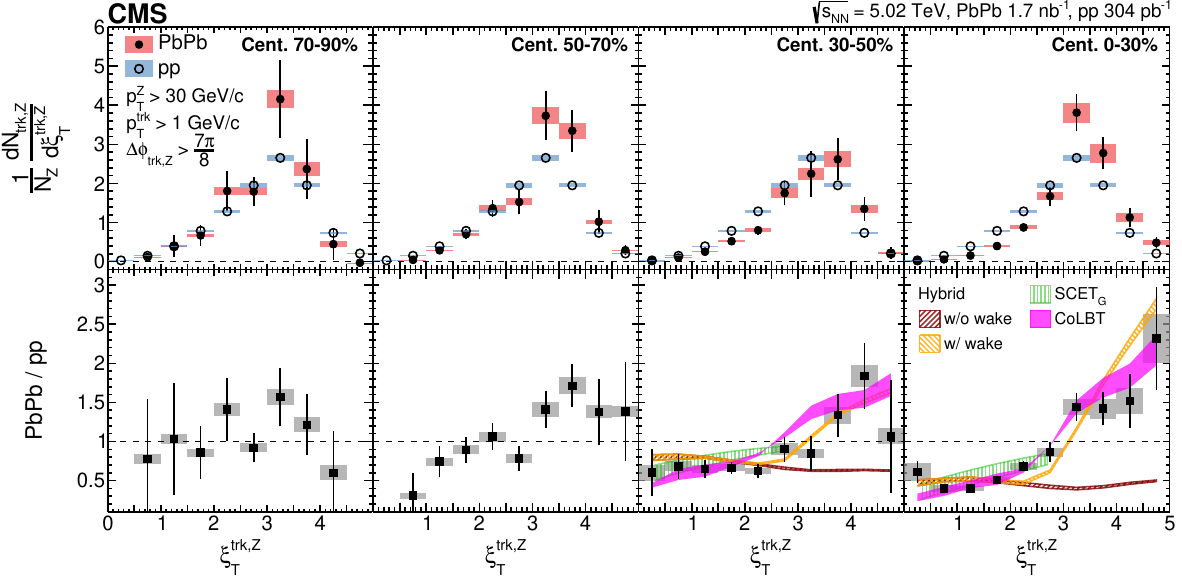}   
      \caption{Upper: distributions of \xizt in \pp collisions compared to \PbPb collisions (left to right) in the 70--90 (left), 50--70, 30--50, and 0--30\% (right) centrality intervals. Lower: ratios of the \PbPb to \pp distributions. The vertical bars and shaded boxes represent the statistical and systematic uncertainties, respectively. Several model calculations are added for comparison: Hybrid~\cite{Casalderrey-Solana:2016jvj}, CoLBT~\cite{Chen:2020tbl,Yang:2021qtl}, and \scetg~\cite{Li:2019dre}.}
      \label{fig:xizh_data}
\end{figure*}

The fragmentation function of the parton emitted back-to-back with the \PZ boson is studied via the \dNxizt distributions, shown in Fig.~\ref{fig:xizh_data}. For these results (as well as for those shown in Fig.~\suppRef, tracks are required to satisfy $\dphitz > 7\pi/8$. Because the interest is in the shape dissimilarities, the ratios of the \pp and \PbPb distributions are presented. All distributions are normalized by the number of \PZ candidates found in each data set. 

In Fig.~\ref{fig:xizh_data}, the low- and high-\xizt regions (\ie, below and above $\sim$3) correspond to high- and low-\pt particles (or lower- and higher-\ptz), respectively. For instance, for $\ptz\sim30(60)\GeVc$, the high-\xizt region corresponds to $\ptt\sim1.5(3)\GeVc$. No significant modification is observed in the 70--90\% PbPb collisions compared to the \pp data. In central collisions, charged particles are suppressed in the $\xizt < 3$ (high-energy particles) interval, and enhanced in the $\xizt > 3$ interval. These features are consistent with a scenario in which the initial parton loses energy (\ie, jet quenching) and the medium induces modification of the parton shower. The enhancement is also consistent with a picture in which additional low-energy particles are produced from the recoil of the medium caused by the traversing parton.

To confirm the onset of medium-induced effects and further help pinpoint the transition point in momentum space for different parton-medium interactions, a comparison of the per-\PZ-boson associated yields in \PbPb and \pp collisions (\dNptt) is needed. Figure~\suppRef shows such a comparison, together with the ratio of the \PbPb and \pp distributions.
In the most peripheral event class, there is no significant modification of the charged-particle \pt spectrum in \PbPb collisions. In central events and at high \ptt ($>$2--5\GeVc), the particle production is suppressed in \PbPb compared to the \pp reference data. At the same time, at low \ptt (1--2\GeVc), an enhancement is observed consistent with the one seen in the \dphitz results. Modifications of the $\xizt$ and $\ptt$ distributions are the largest in the 0--30\% centrality interval, indicating the strongest medium effects. Qualitatively similar observations were reported in photon-~\cite{Sirunyan:2018qec,ATLAS:2019dsv} and Z-tagged~\cite{ATLAS:2020wmg} measurements.

The medium response is not expected to play an important role for the high-\ptt and low-\xizt regions, as is illustrated in the Hybrid model, where calculations with and without wake are indistinguishable. In this region, there is good agreement between the data and the \scetg and the Hybrid calculations. At low-\ptt and high-\xizt, the increase in the charged particle yield can only be reproduced if a feedback from the medium is considered. In these regions, both the Hybrid with wake and CoLBT models capture the general features seen in data, including the expected weakening of medium effects at higher \pt values from 0--30 to 30--50\% \PbPb event centralities.

In summary, the measurements of charged hadrons produced in the shower of a parton originating in the same hard scattering with a \PZ boson, are reported in lead-lead nuclei (\PbPb) and proton-proton (\pp) collisions at $\sqrtsNN = 5.02\TeV$. Collision data with a \PZ boson candidate with transverse momentum $\pt > 30\GeVc$ are analyzed. The \PZ-tagged fragmentation functions and \ptt spectra, which probe the longitudinal structure of the parton shower inside the medium, are measured and significant modifications are observed. Particle yields, which are sensitive to modification of the in-medium parton shower and medium recoils, are measured for all charged particles as a function of the azimuthal angle ($\phi$) with respect to the \PZ boson momentum vector. Comparison of the \PbPb and \pp results indicates a modification of the angular correlation functions extending to $\phi$ angles close to the \PZ boson in central PbPb events. The data favor theoretical models that include the response of the medium to the traversing parton in addition to energy loss. These results represent the first studies of parton-medium interactions over all $\phi$ angles, in which the initial state of the scattered parton is known before it enters the medium. 

\begin{acknowledgments}
   We congratulate our colleagues in the CERN accelerator departments for the excellent performance of the LHC and thank the technical and administrative staffs at CERN and at other CMS institutes for their contributions to the success of the CMS effort. In addition, we gratefully acknowledge the computing centers and personnel of the Worldwide LHC Computing Grid and other centers for delivering so effectively the computing infrastructure essential to our analyses. Finally, we acknowledge the enduring support for the construction and operation of the LHC, the CMS detector, and the supporting computing infrastructure provided by the following funding agencies: BMBWF and FWF (Austria); FNRS and FWO (Belgium); CNPq, CAPES, FAPERJ, FAPERGS, and FAPESP (Brazil); MES (Bulgaria); CERN; CAS, MoST, and NSFC (China); COLCIENCIAS (Colombia); MSES and CSF (Croatia); RIF (Cyprus); SENESCYT (Ecuador); MoER, ERC PUT and ERDF (Estonia); Academy of Finland, MEC, and HIP (Finland); CEA and CNRS/IN2P3 (France); BMBF, DFG, and HGF (Germany); GSRT (Greece); NKFIA (Hungary); DAE and DST (India); IPM (Iran); SFI (Ireland); INFN (Italy); MSIP and NRF (Republic of Korea); MES (Latvia); LAS (Lithuania); MOE and UM (Malaysia); BUAP, CINVESTAV, CONACYT, LNS, SEP, and UASLP-FAI (Mexico); MOS (Montenegro); MBIE (New Zealand); PAEC (Pakistan); MSHE and NSC (Poland); FCT (Portugal); JINR (Dubna); MON, RosAtom, RAS, RFBR, and NRC KI (Russia); MESTD (Serbia); SEIDI, CPAN, PCTI, and FEDER (Spain); MOSTR (Sri Lanka); Swiss Funding Agencies (Switzerland); MST (Taipei); ThEPCenter, IPST, STAR, and NSTDA (Thailand); TUBITAK and TAEK (Turkey); NASU (Ukraine); STFC (United Kingdom); DOE and NSF (USA).
\end{acknowledgments}

\bibliography{auto_generated}  

\ifthenelse{\boolean{cms@external}}{}{
\clearpage
\appendix
\numberwithin{table}{section}
\numberwithin{figure}{section}
\section{Supplemental material\label{app:suppMat}}
\begin{table*}[htb]
\centering
\topcaption{Relative systematic uncertainties for \dphitz, \xizt, and \ptt, averaged over the whole distribution, for \pp and \PbPb collisions. The relative uncertainties were calculated separately for each \dphitz, \xizt, and \ptt bin, and then the quoted average was calculated assuming each bin has the same weight within individual distributions.}

\begin{scotch}{l *2c *2c *2c}
\multirow{2}{*}{Systematic uncertainty source}  & \multicolumn{2}{c}{\dphitz [\%]} & \multicolumn{2}{c}{\xizt [\%]} & \multicolumn{2}{c}{\ptt [\%]} \\
                                                                            & \pp & \PbPb                       & \pp & \PbPb                   & \pp & \PbPb \\
\hline

Tracking efficiency   &  &  &  & & &  \\
Data--MC difference                                           & 2.4 & 5.0 & 2.4 & 5.0 & 2.4 & 5.0 \\
MC minbias--embedding samples difference     & \NA   & 2.7 & \NA     & 2.7 & \NA     & 2.5 \\
MC generator--reconstruction difference            & 0.7 & 8.1 & 2.3 & 5.0 & 1.7 & 3.8 \\

  &  &  &  & & &  \\
Leptons  &  &  &  & & &  \\
Data--MC efficiency difference & 0.4 & 0.7 & 0.4 & 0.7 & 0.4 & 0.7 \\
Energy scale                             & 0.1 & 1.7 & 0.4 & 2.6 & 0.2 & 1.7 \\

  &  &  &  & & &  \\
Pileup \pp                & 1.3 & \NA & 1.9 & \NA & 2.0 &  \NA \\
Event mixing \PbPb & \NA    & 3.0 & \NA & 1.8 & \NA & 0.8 \\

\end{scotch}

\label{tab:systUncert}
\end{table*}

\begin{figure*}[hbt!]
   \centering
      \includegraphics[width=1.0\textwidth]{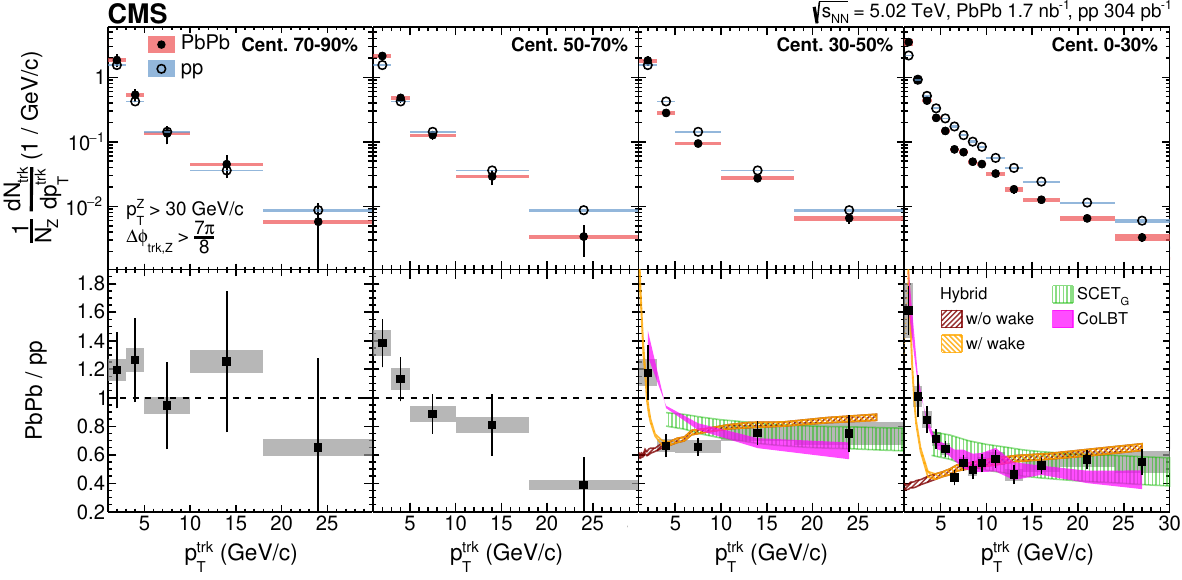}      
      \caption{Upper: Distributions of \dNptt in \pp collisions compared to \PbPb collisions (left to right) in the 70--90 (left), 50--70, 30--50, and 0--30\% (right) centrality intervals. Lower: ratios of the \PbPb to \pp distributions. The vertical bars and shaded boxes represent the statistical and systematic uncertainties, respectively. Several model calculations are added for comparison: Hybrid~\cite{Casalderrey-Solana:2016jvj}, CoLBT~\cite{Chen:2020tbl,Yang:2021qtl}, and \scetg~\cite{Li:2019dre}.}
      \label{fig:trkpt_data}
\end{figure*}
}

\cleardoublepage \section{The CMS Collaboration \label{app:collab}}\begin{sloppypar}\hyphenpenalty=5000\widowpenalty=500\clubpenalty=5000\vskip\cmsinstskip
\textbf{Yerevan Physics Institute, Yerevan, Armenia}\\*[0pt]
A.M.~Sirunyan$^{\textrm{\dag}}$, A.~Tumasyan
\vskip\cmsinstskip
\textbf{Institut f\"{u}r Hochenergiephysik, Wien, Austria}\\*[0pt]
W.~Adam, F.~Ambrogi, T.~Bergauer, M.~Dragicevic, J.~Er\"{o}, A.~Escalante~Del~Valle, R.~Fr\"{u}hwirth\cmsAuthorMark{1}, M.~Jeitler\cmsAuthorMark{1}, N.~Krammer, L.~Lechner, D.~Liko, T.~Madlener, I.~Mikulec, F.M.~Pitters, N.~Rad, J.~Schieck\cmsAuthorMark{1}, R.~Sch\"{o}fbeck, M.~Spanring, S.~Templ, W.~Waltenberger, C.-E.~Wulz\cmsAuthorMark{1}, M.~Zarucki
% ...
\vskip\cmsinstskip
\dag: Deceased\\
1:  Also at Vienna University of Technology, Vienna, Austria\\
% ...
\end{sloppypar}
\end{document}